\documentclass[%
reprint,
showkeys,
 amsmath,amssymb,
 aps,prl
]{revtex4-2}

\usepackage{hyperref}
\hypersetup{
  colorlinks=true,
  linkcolor=blue,
  citecolor=blue,
  urlcolor=blue
}
\usepackage{soul}
\usepackage{cleveref}
\usepackage{siunitx}
\usepackage{miller}
\usepackage{graphicx}
\usepackage{dcolumn}
\usepackage{bm}
\usepackage{graphicx}
\usepackage{tikz}
\usetikzlibrary{calc}
\usetikzlibrary{positioning}
\usepackage{comment}
\usepackage{subfig}
\usepackage{ragged2e}
\usepackage[justification=justified,singlelinecheck=false]{caption}
\newcommand{\nikhil}[1]{\textcolor{red}{Nikhil: \MakeUppercase{#1}}}

\begin{document}

\preprint{APS/123-QED}


\title{
Grain Boundary Phase Transitions Enable Diffusionless Climb of Disconnections 
}

\author{Md Sharier Nazim\textsuperscript{1}}
\author{Giacomo Po\textsuperscript{2}}
\author{Nikhil Chandra Admal\textsuperscript{1}}
\email[Corresponding author: ]{admal@illinois.edu}

\affiliation{\textsuperscript{1}Department of Mechanical Science and Engineering,
University of Illinois , Urbana-Champaign,  Urbana, 61801, IL, USA}

\affiliation{\textsuperscript{2}Department of Mechanical and Aerospace Engineering,
University of Miami, Coral Gables, FL 33124, USA}

\begin{abstract}

Dislocation–grain boundary (GB) interactions govern the mechanical response of polycrystalline materials by controlling dislocation absorption, transmission, and interfacial plasticity. While disconnection climb is conventionally understood to require the absorption or emission of point defects supplied through long-range bulk diffusion, GBs possess intrinsic configurational degrees of freedom associated with their atomic structure, or microstate, that may provide an alternative mechanism. Here, using bicrystallography and molecular dynamics simulations, we investigate the interaction of shear dislocation loops with the $\hkl[110]\hkl(-5 5 14)$  symmetric tilt grain boundary in Al. We show that dislocation absorption generates a mobile extrinsic disconnection with a nonzero climb component that propagates conservatively along the interface without long-range bulk point-defect transport. Its motion is accompanied by a localized GB phase transformation mediated by cooperative atomic rearrangements within the GB core, producing successive metastable GB microstates. These findings establish a direct coupling between lattice dislocations and GB phase evolution and reveal a conservative mechanism for disconnection climb fundamentally distinct from conventional vacancy-mediated climb.

\end{abstract}



\maketitle

Plastic deformation and microstructural evolution in crystalline materials are governed by the dynamics of line defects. In bulk crystals, plasticity is mediated by dislocations, whereas grain boundaries (GBs) support a broader class of interfacial line defects known as disconnections \cite{han2018grain,anderson2017theory,dimitrakopulos1997defect}. Dislocations exist within the crystal lattice and preserve lattice connectivity across a glide plane, while disconnections reside on interfaces and additionally possess a step character associated with changes in GB morphology \cite{hirth1996steps,sutton1995interfaces}. Disconnections play a central role in GB migration \cite{chen2020temperature,chen2020mobilitytensor}, microstructural evolution \cite{han2018grain,hirth1996steps,babcock1989grain1,babcock1989grain}, sliding,  and dislocation transmission across interfaces \cite{han2018grain,cahn2006coupling,cahn2004unified}. Structural phase transformations at grain boundaries \cite{frolov2013structural, devulapalli2024topological,meiners2020observations,langenohl2022dual,brink2023universality,chen2020grain} have emerged as a fundamental aspect of interface behavior, influencing grain boundary migration, defect interactions, and the mechanical response of crystalline materials \cite{dillon2007complexion,cantwell2014grain,dillon2026basic,cantwell2020grain}. Recent developments in bicrystallography and interface defect theory have further established that GB microstates themselves can be interpreted in terms of intrinsic disconnection networks and their interactions \cite{admal2022interface,joshi2026equilibrium,deka2023automated}.

Dislocations and disconnections share important geometric similarities. Both are line defects characterized by a Burgers vector and a local line direction, and both may undergo glide or climb motion \cite{hirth2006disconnections,hirth1996steps,hirth2013interface}. Glide corresponds to motion in which the defect velocity lies in the plane spanned by the Burgers vector and the line tangent, whereas climb corresponds to motion outside this plane \cite{hull2011introduction,anderson2017theory}.

In bulk crystals, dislocation climb is mediated by the absorption or emission of vacancies and interstitials, which redistribute mass through long-range diffusion \cite{hull2011introduction,anderson2017theory,kabir2010predicting,thompson2018interstitial}. The rate of distortion ($\dot{\bm \beta}^{\rm d}$) associated with diffusion compensates for the rate of volumetric plastic distortion ($\dot{\bm\beta}^{\rm p}$) generated by dislocation climb, so that the total inelastic deformation occurs without local density changes. In other words,
\begin{equation}
\mathrm{tr}(\dot{\bm \beta}^{\rm p})+\mathrm{tr}(\dot{\bm \beta}^{\rm d})=0.
\label{eq:dislocation_climb_balance}
\end{equation}
Because diffusion is thermally activated, dislocation climb becomes important primarily during high-temperature deformation \cite{hull2011introduction}.

This physical picture has strongly influenced the interpretation of disconnection climb, leading to the widespread assumption that non-conservative GB plasticity must likewise be accommodated through point-defect diffusion and mass exchange with neighboring grains \cite{han2018grain,barr2022irradiation,magri2020coupled,rajabzadeh2013evidence,rajabzadeh2014role,mcelfresh2021discrete}. Recent atomic-resolution observations of disconnection climb in ceramic GBs have further reinforced this interpretation by directly associating GB migration with vacancy transport \cite{wei2021direct,wei2022direct,feng2023atomistic}.

In this Letter, we show that the compensating mechanism for disconnection climb need not involve long-range bulk diffusion. Unlike bulk crystals, GBs possess an internal configurational space composed of distinct microstates with different excess densities and defect populations \cite{frolov2013structural,admal2022interface,joshi2026equilibrium}. As a result, the local density changes associated with disconnection climb can be accommodated by structural phase transformations within the GB itself, without the need for long-range transport of point defects through the adjoining grains. In this case, the volumetric plastic strain, $\bm \beta^{\rm p}$, generated by disconnection climb is compensated by the transformation strain, $\bm \beta^{\rm t}$ associated with the evolving GB structure,
\begin{equation}
\mathrm{tr}(\dot{\bm \beta}^{\rm p})+\mathrm{tr}(\dot{\bm \beta}^{\rm t})=0.
\label{eq:disconnection_climb_balance}
\end{equation}
This mechanism plays a role analogous to diffusion during dislocation climb, but does not rely on thermally activated bulk transport.Consequently, disconnection climb need not be restricted to high temperatures. This finding suggests that non-conservative GB plasticity may remain operative under conditions where diffusion-mediated climb of lattice dislocations is effectively frozen, with important implications for GB migration, sliding, and structural evolution in crystalline materials. More broadly, it identifies a previously unrecognized low-temperature mechanism for non-conservative GB plasticity.

\begin{figure}[t!]
    \centering
    \includegraphics[scale=0.4]{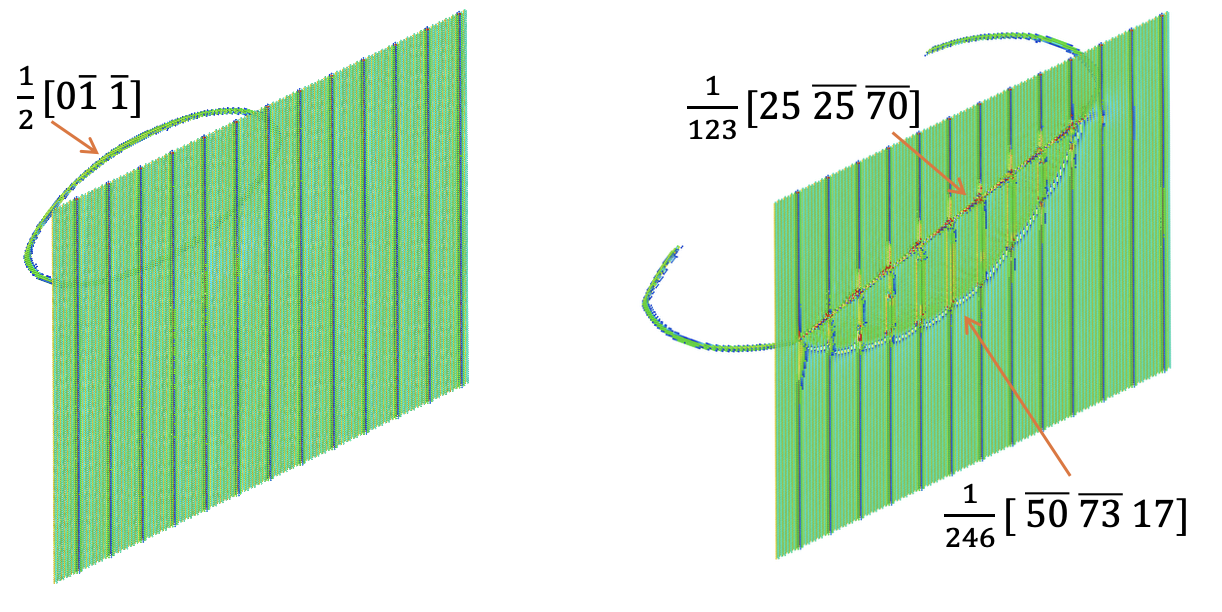}
    \caption{\justifying Centrosymmetry analysis (OVITO \cite{stukowski2009visualization}) of the GB plane shows dislocation absorption facilitated by GB plasticity.}
    \label{fig:cs}
\end{figure}

\begin{figure*}[t]
    \centering
    \includegraphics[scale=0.35]{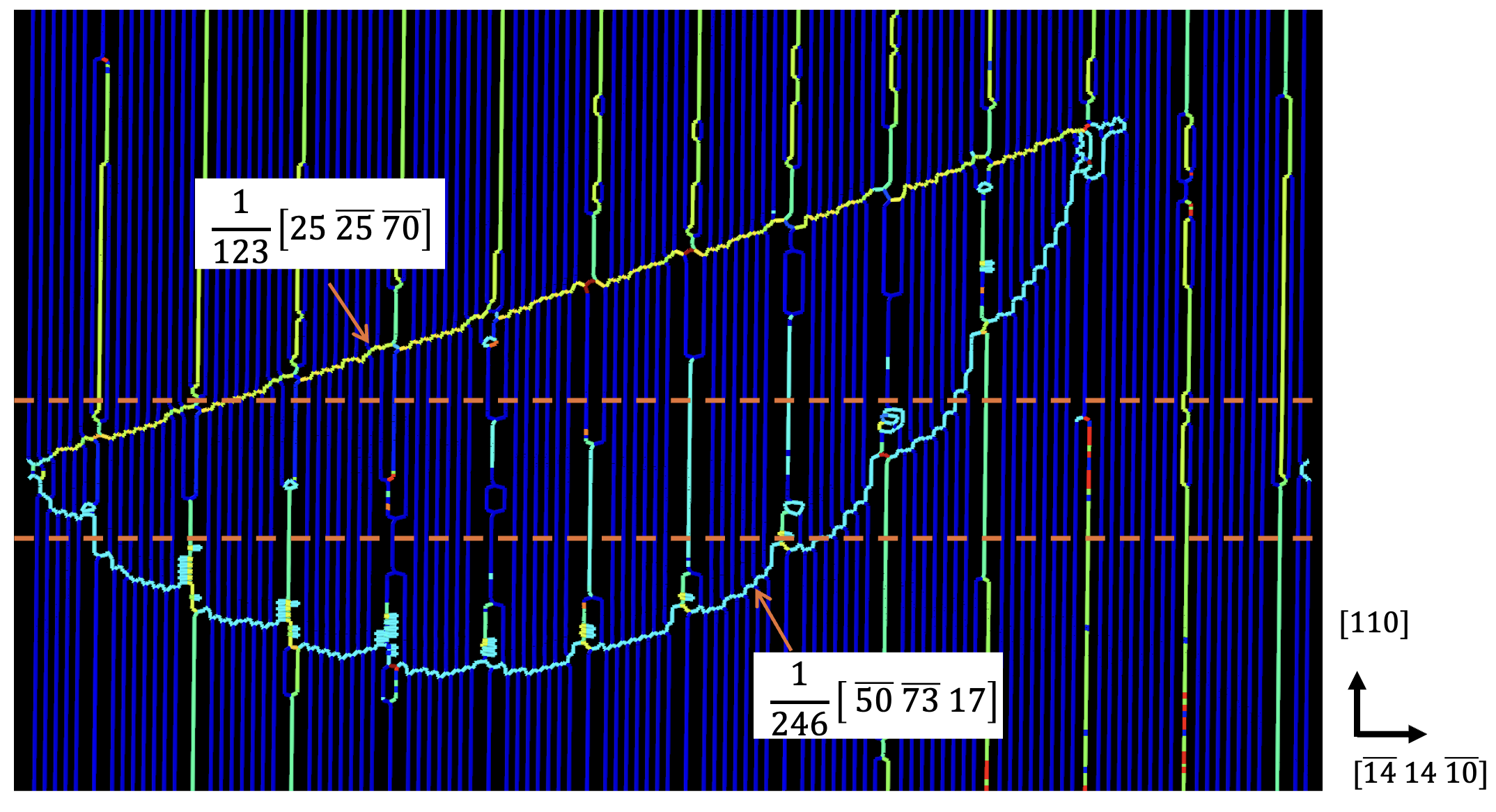}
    \includegraphics[scale=0.35]{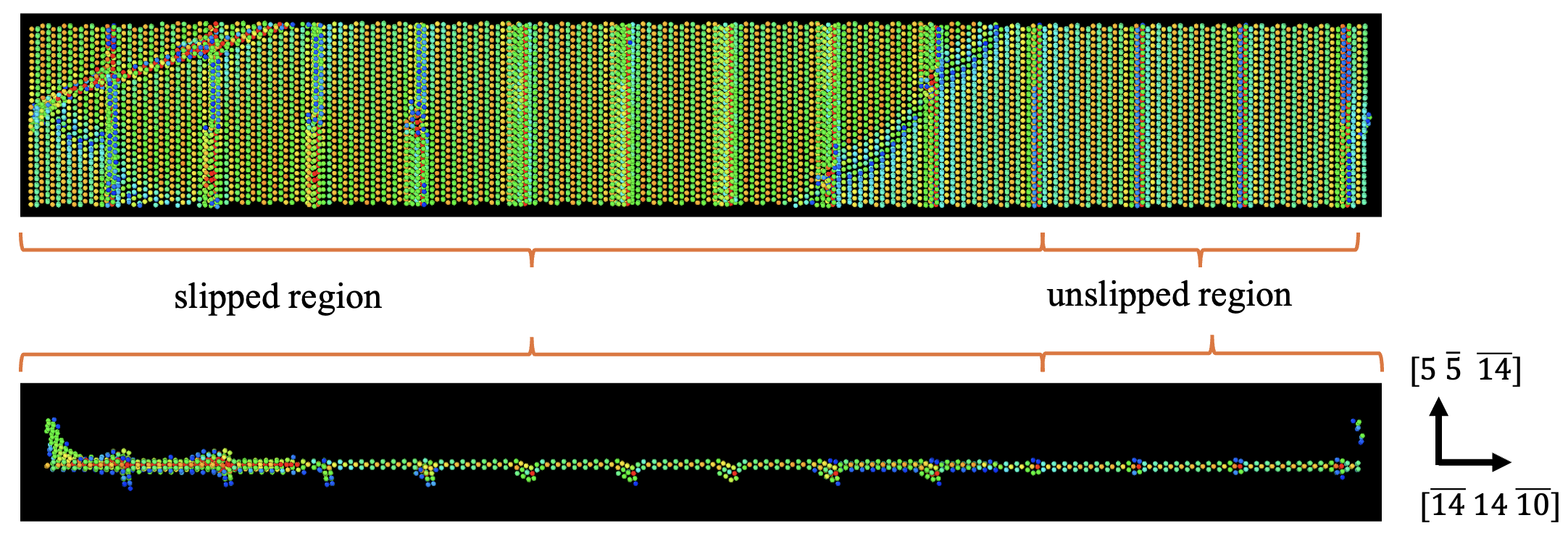}

    \caption{\justifying A cross-sectional view of the GB illustrates structural changes associated with GB phase transitions. The dislocation dissociates into two extrinsic disconnections: a sessile (yellow) disconnection, which is immobile, and a mobile (cyan) disconnection. Edge-on and top views of the GB during dislocation impingement demonstrate that the region swept by the mobile disconnection has slipped and undergone a phase transition, while the unslipped region remains in its minimized, stable phase.}
    \label{fig:csmnew}
\end{figure*}

Using atomistic simulations, we demonstrate that a climb disconnection with a nonzero Burgers-vector component normal to the interface can propagate along a GB without long-range diffusion through cooperative atomic rearrangements within the GB core. Rather than being accommodated by vacancy-mediated mass transport through the adjoining grains, the resulting motion is accompanied by localized transformations of the GB microstructure, i.e., GB phase transitions.

To demonstrate this mechanism, we intentionally avoided constructing a highly specialized simulation geometry designed to enforce conservative disconnection climb. Instead, climb disconnections were allowed to emerge naturally during a generic dislocation--GB interaction. Molecular dynamics (MD) simulations were therefore performed using the Large-scale Atomic/Molecular Massively Parallel Simulator (LAMMPS) \cite{thompson2022lammps} on a $\Sigma123\, \hkl[110]\hkl(-5\,5\,14)$ symmetric tilt grain boundary (STGB) in Al \cite{bamney2022disclination}, modeled using the embedded-atom-method potential of Mishin \textit{et al.} \cite{mishin1999interatomic}. The bicrystal was constructed using the Smith normal form bicrystallography framework of \citet{admal2022interface}.

The GB plane was oriented normal to the $x$-axis, with the tilt axis aligned with $z$. Periodic boundary conditions were applied in all directions, with vacuum layers introduced along the GB normal direction. Following the statistical-mechanical framework of \citet{joshi2026equilibrium}, which interprets a GB microstate as a network of intrinsic disconnections, we systematically enumerated and constructed admissible GB microstates. This procedure identified the lowest-energy microstate, which adopts a characteristic diamond-shaped structural motif \cite{bamney2022disclination} following energy minimization using the fast inertial relaxation engine (FIRE) algorithm.

A shear dislocation loop of radius $37a_0$ was subsequently introduced into one grain using its analytical elastic displacement field \cite{po2014recent}. A resolved shear stress of $\tau=\SI{1200}{\mega\pascal}$ was then applied to drive the loop toward the GB. Specifically, the applied stress tensor was
\begin{equation}
\sigma=\tau\left(\hat{\boldsymbol{s}}\otimes\hat{\boldsymbol{n}}+\hat{\boldsymbol{n}}\otimes\hat{\boldsymbol{s}}\right),
\end{equation}
where $\hat{\boldsymbol{s}}$ and $\hat{\boldsymbol{n}}$ denote the unit slip direction and slip-plane normal, respectively. This loading configuration was chosen to suppress cross-slip. The stress was gradually increased to the target value at a constant strain rate, during which the full dislocation dissociated into partials. The system was then allowed to evolve as an $(NVT)$-ensemble with $T=\SI{2}{\kelvin}$.

Under the applied shear stress, the expanding lattice dislocation impinges on the GB and dissociates into a mobile and a sessile residual disconnection,\footnote{Although this reaction is not favored by the Burgers content alone
($|\bm b_{\rm mobile}|^2+|\bm b_{\rm sessile}|^2 \approx 0.54\,a_0^2 > |\bm b_{\rm disl}|^2 = 0.50\,a_0^2$),
it is driven by the elimination of the impinging lattice dislocation at the boundary and by the
work of the applied stress.}
\begin{equation}
\underbrace{\frac{1}{2}\hkl[0\, -1\, -1]}_{\text{dislocation}}
=
\underbrace{\frac{1}{246}\hkl[-50\, -73\, 17]}_{\text{mobile disconnection}}
+
\underbrace{\frac{1}{123}\hkl[25\, -25\, -70]}_{\text{sessile disconnection}}.
\end{equation}

\Cref{fig:cs} shows centrosymmetry maps of the interaction. The sessile disconnection remains localized near the point where the incoming dislocation intersects the GB, whereas the mobile disconnection expands as a loop within the interface. The corresponding Burgers vector-step height pairs of the disconnection modes are $(b,h) = (2.58,0)\,\si{\angstrom}$ and $(1.48,1.23)\,\si{\angstrom}$, respectively. 
The absorption partitions the impinging dislocation's large normal-Burgers content between an immobile, purely prismatic residual---the sessile disconnection, whose Burgers vector is exactly antiparallel to the GB normal---and the mobile disconnection, which carries only a small climb component ($\bm b\cdot\hat{\bm n}\approx\SI{0.13}{\angstrom}$, $\approx 9\%$ of $|\bm b|$). It is the
conservative propagation of this small but nonzero climb component that requires accommodation.




According to the conventional picture of disconnection climb, propagation of a disconnection with a nonzero normal Burgers-vector component requires the absorption or emission of point defects supplied through long-range diffusion from the neighboring grains \cite{han2018grain}. The present simulations reveal a qualitatively different mechanism. The mobile disconnection propagates conservatively along the interface while remaining confined to the GB, without any evidence of long-range mass transport through the adjoining crystals. Because the simulation is performed at $\SI{2}{\kelvin}$, thermally activated bulk diffusion is entirely frozen; the climb therefore proceeds in a regime where the conventional vacancy-mediated mechanism is categorically unavailable, isolating the GB-intrinsic accommodation.

Prior to dislocation absorption, the GB resides in its lowest-energy microstate. The corresponding intrinsic disconnection network, identified using the Interface Line Defects Analysis (ILDA) framework \cite{deka2023automated}, is shown in the top panel of \Cref{fig:csmnew}. As the mobile disconnection traverses the interface, the atomic structure of the GB undergoes a localized transformation within the swept region, while the unswept portion of the boundary remains in its original state. This transformation is evident in both the intrinsic disconnection networks and the corresponding centrosymmetry maps shown in \Cref{fig:csmnew}. Thus, the passage of the mobile disconnection converts one GB microstate into another through a topological reorganization of the intrinsic disconnection network that defines the interface structure. The propagation of the mobile disconnection is therefore intrinsically coupled to a localized GB phase transformation.
For a constant-mass system, conventional diffusion-mediated climb of this disconnection would remove material and contract the bicrystal along the GB normal by \SI{0.1444}{\angstrom} (set by the disconnection's normal Burgers component). We instead measure a net change in height of only \SI{0.04}{\angstrom} across a control volume spanning the boundary---a reference-independent dilatation that requires no definition of ``GB atoms''. Its smallness implies that the climb-induced volume change is compensated internally by a transformation distortion $\bm\beta^{\rm t}$, which we thus infer rather than measure directly.


\begin{figure*}[t]
    \centering

    \begin{minipage}[c]{0.45\textwidth}
        \centering

        \includegraphics[width=\linewidth]{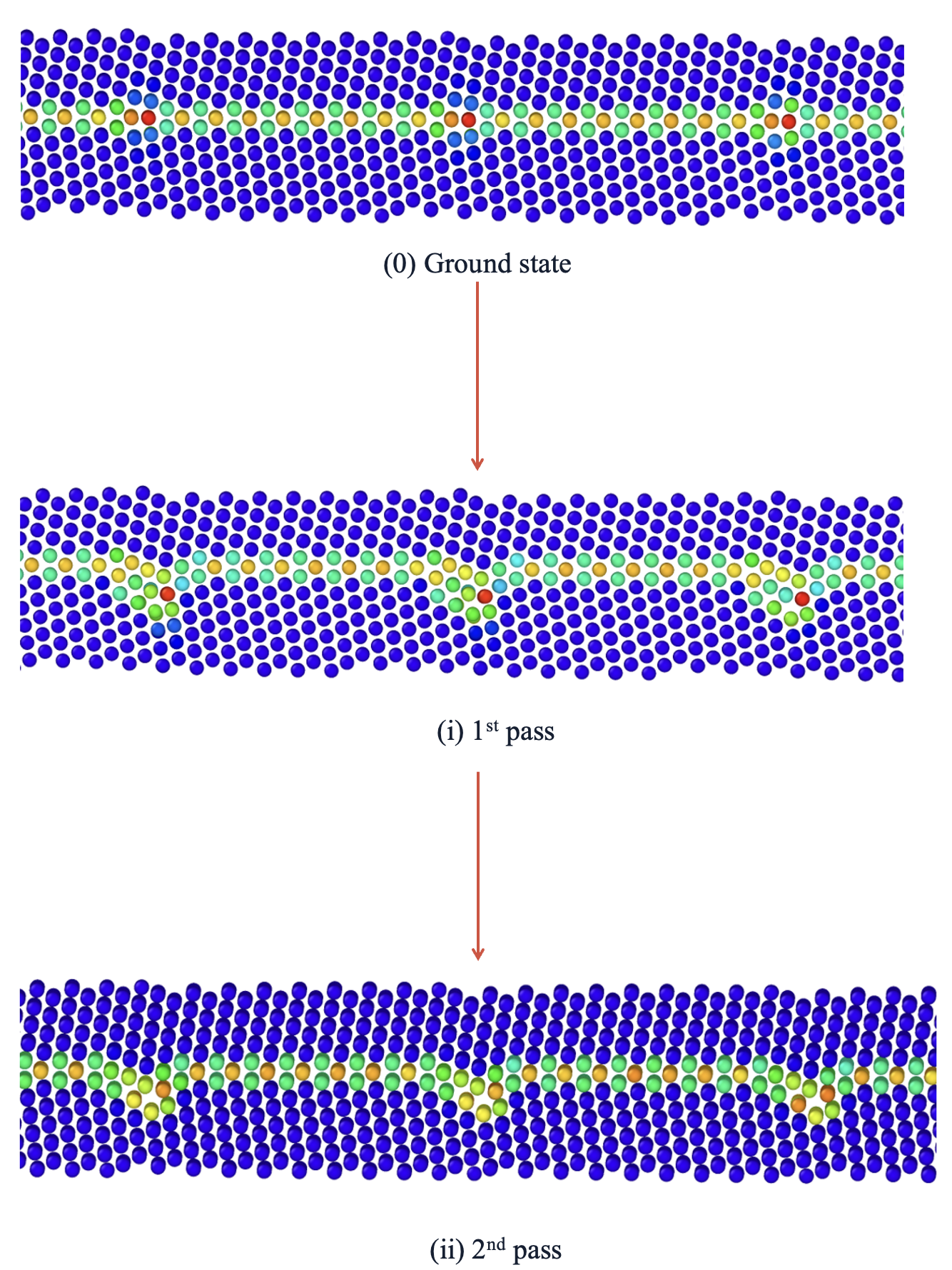}\\[0.4em]

        {\textbf{(A)}}
    \end{minipage}
    \hfill
    \begin{minipage}[c]{0.45\textwidth}
        \centering

        \makebox[\linewidth][c]{%
            \includegraphics[width=1.15\linewidth]{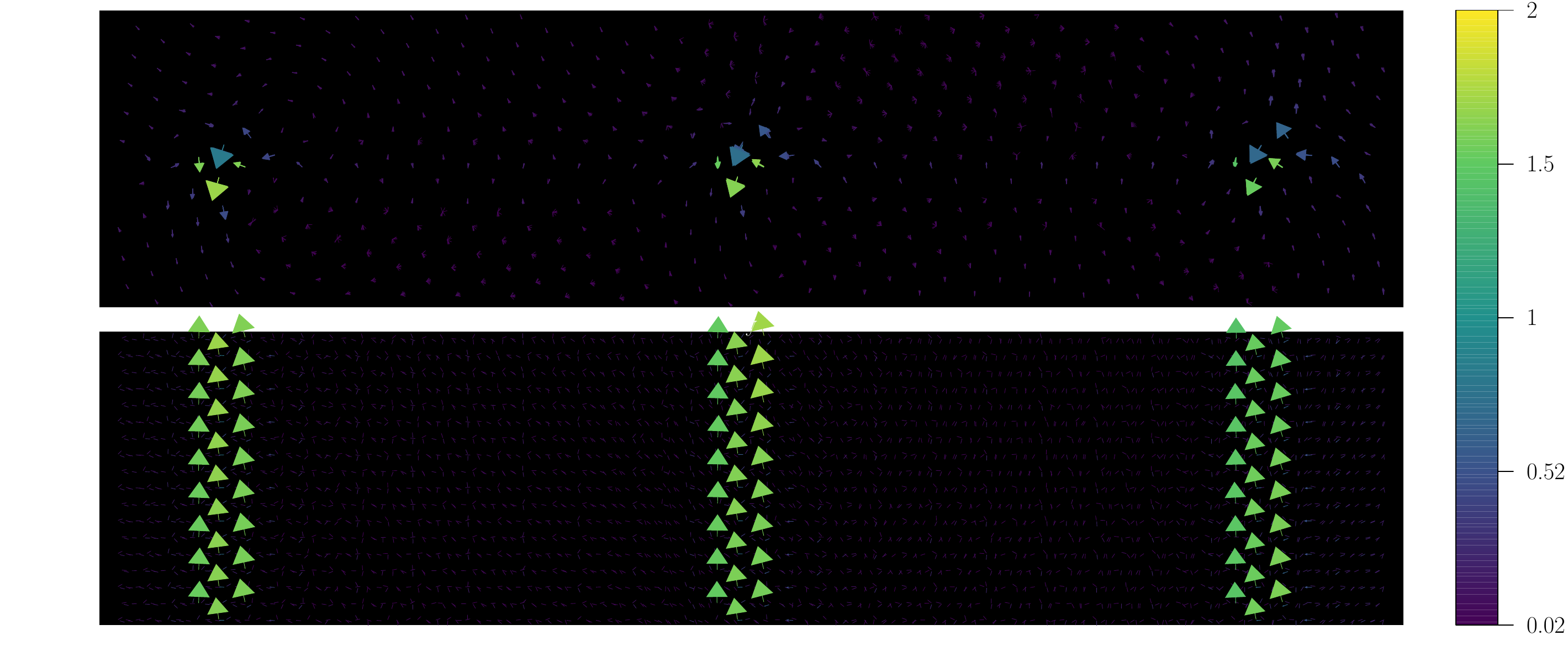}
        }\\[3 em]

        \makebox[\linewidth][c]{%
            \includegraphics[width=1.15\linewidth]{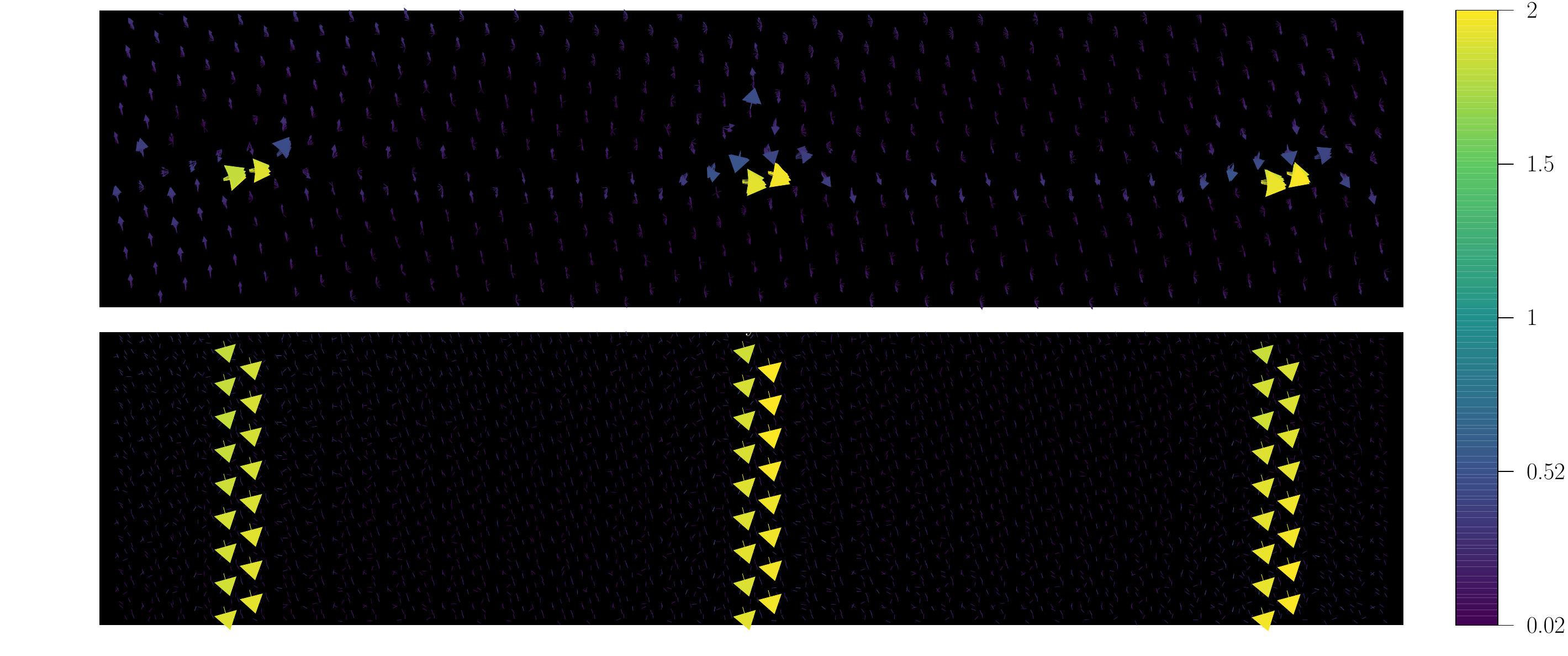}
        }\\[0.4em]

        {\textbf{(B)}}
    \end{minipage}
    \caption{\justifying Successive GB phase transformations induced by repeated passage of the climb disconnection. Panel A shows the GB structures after the 0th (ground state), 1st, and 2nd dislocation passages, illustrating the generation of successive metastable GB microstates.  Panel B presents the corresponding differential displacement fields (edge-on and top views) for the ground-state $\rightarrow$ first and first $\rightarrow$ second transformations. The displacement fields reveal that each phase transformation is accomplished by cooperative atomic shuffling confined to the GB core, generating new GB microstates through localized structural rearrangements without long-range bulk diffusion. Arrows indicate the dominant atomic displacements (in $\si{\angstrom}$) associated with each transformation. 
    }
    \label{fig:two_column_layout}
\end{figure*}


The structural evolution induced by repeated disconnection passage is illustrated in \Cref{fig:two_column_layout}. Owing to the periodic boundary conditions, the region of interest is swept multiple times by the mobile climb disconnection. Each passage transforms the local interface into a distinct metastable GB microstate characterized by a different intrinsic disconnection network and excess density. Differential displacement analysis shown in \Cref{fig:two_column_layout}B reveals that these transformations proceed through cooperative atomic shuffling confined to the GB core. The associated atomic displacements are highly localized and exhibit essentially no net transport normal to the interface, providing no evidence of long-range vacancy-mediated mass transfer through the adjoining grains. Instead, the excess density required to accommodate disconnection climb is supplied by the structural transformation of the GB itself. These observations demonstrate that the succession of GB microstates is generated through localized interfacial rearrangements rather than diffusion-mediated mass transport. 

Our findings have three important consequences.

\begin{figure}[t!]
    \centering
    \subfloat[Disconnection climb with diffusion\label{fig:conservative}]{%
        \includegraphics[scale=0.2]{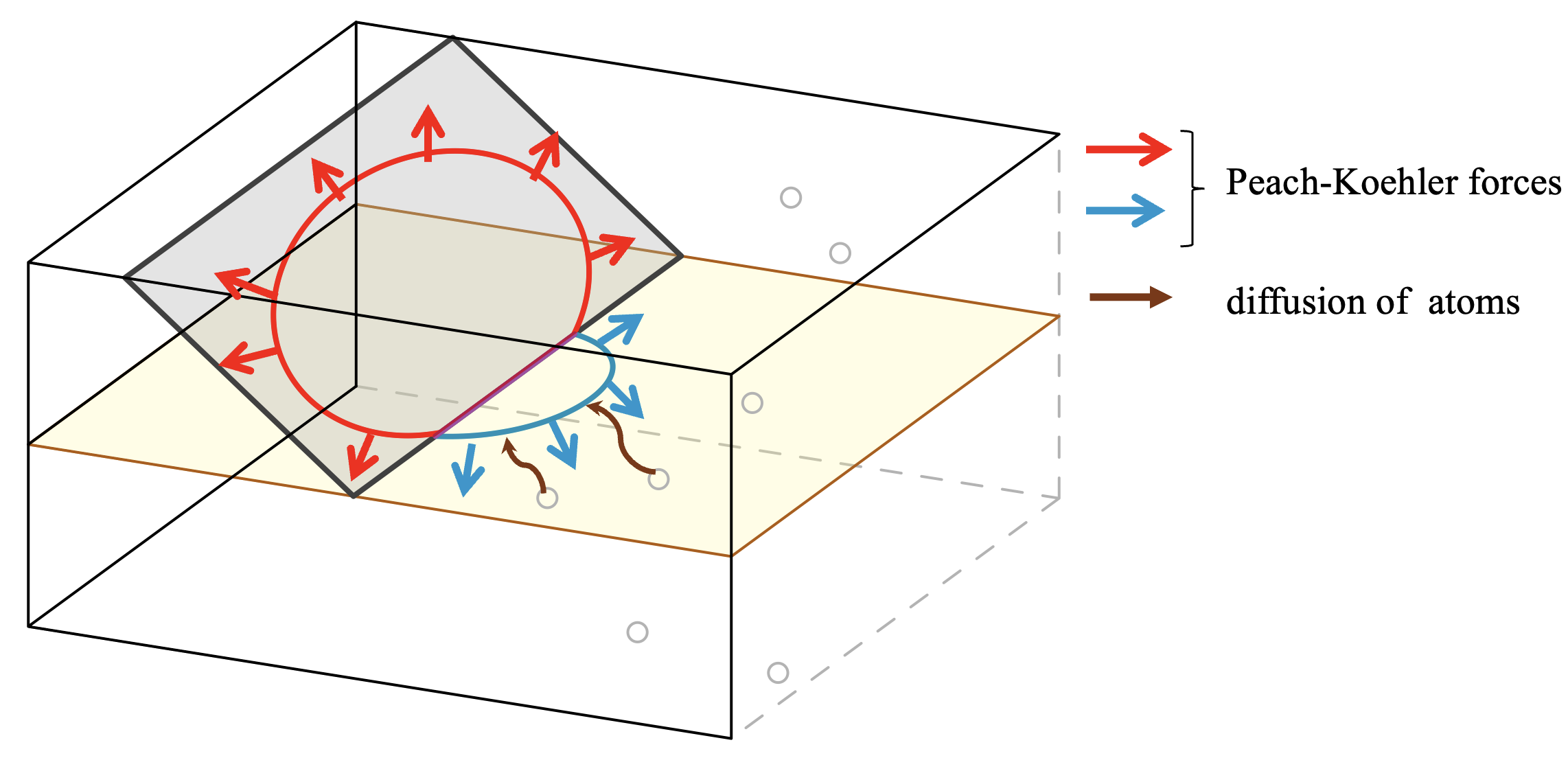}
    }\\
    \subfloat[Disconnection climb with phase transition \label{fig:nonconservative}]{%
        \includegraphics[scale=0.2]{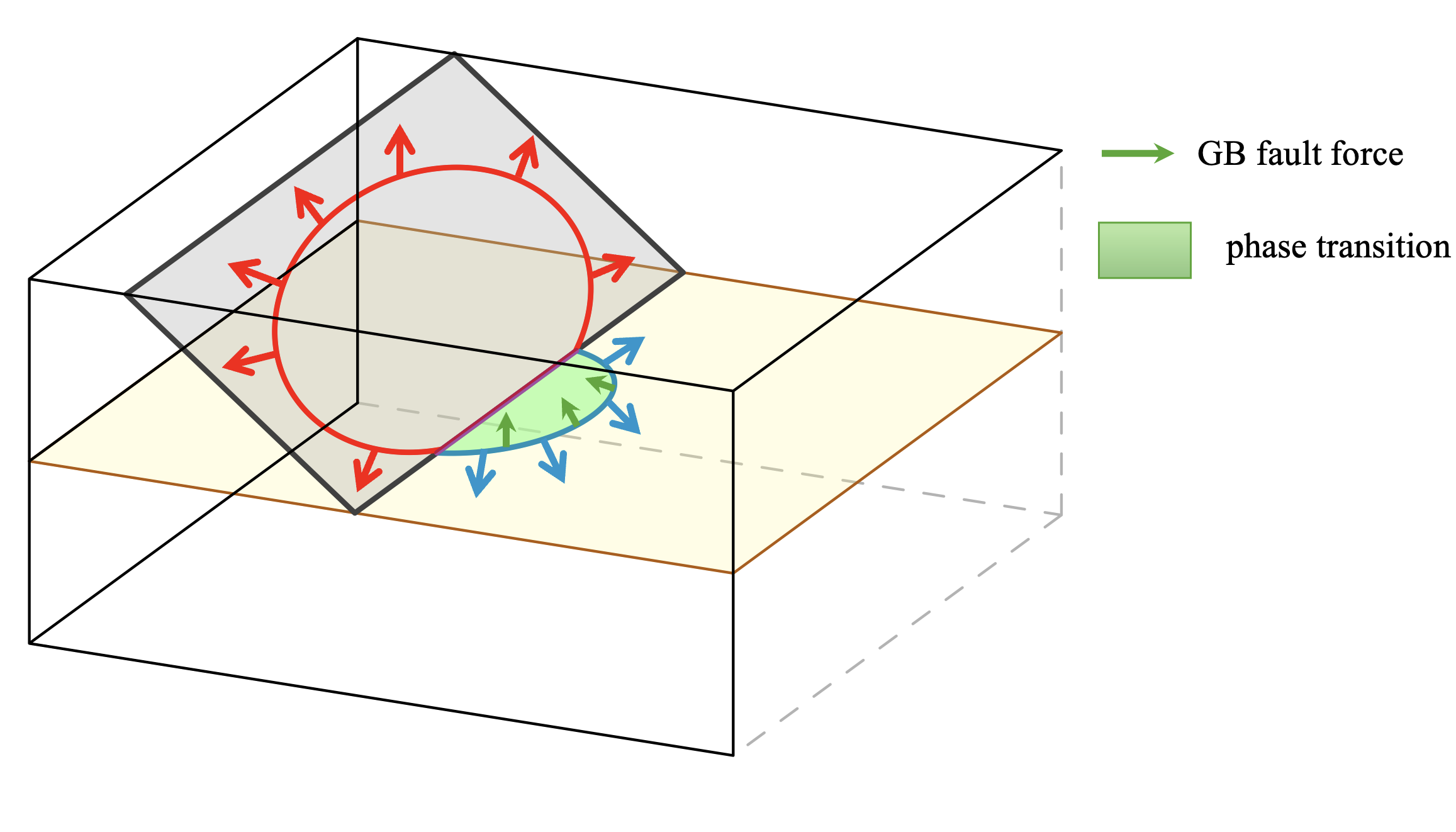}
    }
    \caption{\justifying Two possible modes of climb of a disconnection (blue) nucleated by a bulk dislocation (red). (a) shows climb of the disconnection via the absorption of bulk point defects while the GB structure is preserved. Only mechanical forces are acting on the disconnection in this case (b) shows a disconnection climb mediated by GB phase transition. The area undergoing the phase transformation is shaded in color green, while the green arrows represent configurational chemical forces corresponding to the energy difference between the two GB phases.}
    \label{fig:chemicalForce}
\end{figure}

\begin{itemize}

\item First, climb disconnections emerge naturally as the ``phase-junction defects'' separating neighboring GB phases or microstates of differing excess density observed in the GB literature \cite{frolov2021dislocation,winter2022nucleation}. This interpretation provides a natural explanation for the propagating mobile disconnection observed in the present simulations. Rather than viewing it as a conventional climb disconnection requiring external point-defect transport, it can be regarded as the phase-junction defect separating transformed and untransformed GB microstates. As the defect advances under an applied mechanical driving force, one GB microstate is continuously converted into another, while the density change associated with the climb component is accommodated internally by the evolving GB structure. Disconnection climb, therefore, emerges as the propagation of a GB phase transformation.

\item Second, the mechanically driven phase-transformation mechanism identified here differs fundamentally from previously reported forms of disconnection climb. In irradiated $\alpha$-Al$_2$O$_3$, for example, disconnection climb proceeds through the absorption of vacancies supplied by the surrounding crystal under electron irradiation \cite{wei2021direct,wei2022direct,feng2023atomistic}. Similarly, recent studies of GB phase transformations have interpreted the migration of phase-junction defects as a thermally activated process driven by the free-energy difference between neighboring GB phases \cite{langenohl2022dual,meiners2020observations}. In contrast, the transformation observed here is mechanically induced by dislocation absorption, and the associated climb component is accommodated by structural rearrangements confined to the GB itself rather than by long-range bulk diffusion. Because this mechanism does not rely on thermally activated point-defect transport, it enables disconnection climb at temperatures where diffusion-mediated climb of lattice dislocations is effectively suppressed. More broadly, it reveals that, unlike in bulk crystals where low-temperature plasticity is dominated by glide, both glide and climb disconnections may contribute significantly to GB plasticity even at low temperatures.

\item  Third, our findings imply that the configurational force acting on a climb disconnection contains not only the mechanical Peach--Koehler contribution but also a chemical, or phase-transformation, contribution associated with the free-energy difference between the GB microstates on either side of the defect (\Cref{fig:chemicalForce}). Within this picture, a climb disconnection acts as a phase-junction defect separating two distinct GB states, and its motion is driven not only by applied stresses but also by the thermodynamic tendency of one GB phase to replace another. This chemical driving force is directly analogous to the stacking-fault force acting on partial and superpartial dislocations in bulk crystals, where defect motion is influenced by the free-energy difference between neighboring crystal structures characterized by distinct stacking sequences. The present results therefore suggest a unified description in which climb disconnections are governed by coupled mechanical and phase-transformation driving forces, linking the mechanics of interface defects to the thermodynamics of GB phase transitions. Within this framework, the free-energy difference between neighboring GB microstates gives rise to a chemical driving force that may represent an important and previously underappreciated contribution to disconnection mobility, analogous to the stacking-fault force acting on partial and superpartial dislocations in bulk crystals. 

\vskip 2pt

\end{itemize}

More broadly, our findings establish climb disconnections as phase-junction defects whose motion can be accommodated by GB phase transformations rather than long-range diffusion, fundamentally revising their role in low-temperature GB plasticity and revealing new thermodynamic contributions to the forces and mobilities governing interface-defect dynamics.



Author Contributions: N.C.A and G.P. conceived and supervised the research. M.S.N. conducted 
molecular dynamics simulations, interface defect analysis, and prepared figures. All authors contributed to 
interpretation and manuscript preparation.

\vskip 15 pt
\begin{acknowledgments}
\textit{Acknowledgments}---N.C.A. and M.S.N. would like to acknowledge support from the National Science Foundation (NSF),  CAREER grant MOMS-2239734.
G.P. acknowledges the support of NSF through the CAREER grant number 2340174, and the US Department of Energy, Office of Fusion Energy Sciences, under Awards Number DE-SC0024675 and DE-SC0024401 with the University of Miami.
\end{acknowledgments}
\vskip 10 pt
\textit{Data availability}---The data are available from the authors upon further request.

\bibliographystyle{apsrev4-2}
\bibliography{apsNew}

\end{document}